\documentclass[12pt,a4paper]{article}

\newcommand{\beq}{\begin{equation}}
\newcommand{\eeq}{\end{equation}}
\newcommand{\bea}{\begin{eqnarray}}
\newcommand{\eea}{\end{eqnarray}}

\newcommand{\tope}{\mbox{\scriptsize top}}
\newcommand{\bote}{\mbox{\scriptsize bot}}

\newcommand{\dd}{\partial}

\usepackage{amsmath, amsxtra, amsfonts, amsbsy, amsthm} 
\usepackage{graphicx}

\title{A modified Lin equation for the energy balance in isotropic turbulence}
\author{W. D. McComb, \\
SUPA School of Physics and Astronomy,\\
Peter Guthrie Tait Road, \\
University of Edinburgh,\\
EDINBURGH EH9 3JZ.\\
Email: wdm@ph.ed.ac.uk \\
Website and blog: blogs.ed.ac.uk/physics-of-turbulence/}

\begin{document}
\maketitle
%suppress page number on title page
\thispagestyle{empty}
%start page numbering after title at 1
%\setcounter{page}{0}
%\newpage
\begin{abstract}

At sufficiently large Reynolds numbers, turbulence is expected to
exhibit scale-invariance in an intermediate (`inertial') range of
wavenumbers, as shown by power-law behaviour of the energy spectrum and
also by a constant rate of energy transfer through wavenumber. However,
there is an apparent contradiction between the definition of the energy
flux (\emph{i.e.} the integral of the transfer spectrum) and the observed
behaviour of the transfer spectrum itself. This is because the transfer
spectrum $T(k)$ is invariably found to have a zero-crossing at a single
point (at $k=k_\ast$, say), implying that the corresponding energy flux
cannot have an extended plateau but must instead have a maximum value at
$k=k_\ast$.   This behaviour was formulated as a paradox and resolved by
the introduction of filtered/partitioned transfer spectra, which
exploited the symmetries of the triadic interactions (J. Phys. A: Math.
Theor., 41:75501, 2008). In this paper we consider the more general
implications of that procedure for the spectral energy balance equation,
also known as the \emph{Lin equation}. It is argued that the resulting
modified Lin equations (and their corresponding Navier-Stokes equations)
offer a new starting point for both numerical and theoretical methods,
which may lead to a better understanding of the underlying energy
transfer processes in turbulence. In particular the filtered-partitioned
transfer spectra could provide a basis for a hybrid approach to the
statistical closure problem, with the different spectra being tackled
using different methods.

\end{abstract}

\newpage

\section{Introduction}

We have previously written about the scale-invariance paradox and shown
how it may be resolved by the introduction of filtered-partitioned forms
of the transfer spectra \cite{McComb08}, \cite{McComb14a}. In the present
paper we carry on this work to show how the underlying symmetries of the
triadic interactions in wavenumber space also have implications for any
more general study of the Lin equation. We have remarked elsewhere that
to treat the Lin equation as purely a local energy balance equation is
to be in danger of failing to realize that it is actually a highly
non-local equation which couples all modes together. It is in fact the
basis of the cascade picture of turbulent energy transfer, and it is
important to always bear in mind that the transfer spectrum can be
written as an integral over all wavenumbers of a term containing the
triple-moment. In the present work we will argue that it is desirable to
extend this scrutiny to the filtered-partitioned forms of the transfer
spectrum in order to achieve a fuller understanding of the basic energy
transfer processes.

This paper is organized as follows. We begin by stating the Lin
equations and making some observations about the conventional
interpretation of its role as an energy balance in wavenumber. Next we
remind ourselves about the scale-invariance paradox and how it may be
resolved. Then we move on to discussing the ways in which the Lin
equation can be modified in order to clarify its role.

\section{The Lin equation}

We begin with the (by now) well-known spectral energy balance equation
in its most familiar form, thus:
\begin{equation}
	\left( \frac{\dd}{\dd t} + 2 \nu k^2 \right) E(k,t) = T(k,t),
\label{enbalt}
\end{equation}
where $E(k,t)$ is the energy spectrum, $T(k,t)$ is the energy transfer
spectrum and $\nu$ is the kinematic viscosity. A full derivation and
discussion will be found in the book \cite{McComb14a}. We will also
follow the growing practice of referring to it as the Lin equation.

Now let us integrate each term of (\ref{enbalt}) with respect to
wavenumber, from zero up to some arbitrarily chosen wavenumber $\kappa$:
\beq
\frac{\dd}{\dd t}\int_{0}^{\kappa} dk\, E(k,t)  = 
\int^{\kappa}_{0} dk\, T(k,t)
-2 \nu\int_{0}^{\kappa} dk\, k^2 E(k,t).
\label{fluxbalt1}
\eeq
 The
energy transfer spectrum may be written as
\beq
T(k,t) = \int^{\infty}_{0} dj\, S(k,j;t),
\label{ts}
\eeq
where, as is well known, $S(k,j;t)$ can be expressed in terms of the
triple moment. Its antisymmetry under interchange of $k$ and $j$
guarantees energy conservation in the form:
\beq
\int^{\infty}_{0} dk\, T(k,t) =0.
\label{encon}
\eeq

With some use of the antisymmetry of $S$, along with equation
(\ref{encon}), equation (\ref{fluxbalt1}) may be written as
\beq
\frac{\dd}{\dd t}\int_{0}^{\kappa} dk\, E(k,t)  = 
- \int^{\infty}_{\kappa} dk\,\int^{\kappa}_{0} dj\, S(k,j;t)
-2 \nu\int_{0}^{\kappa} dk\, k^2 E(k,t).
\label{fluxbalt2}
\eeq
In this familiar form, the integral of the transfer term is readily
interpreted as the net flux of energy from wavenumbers less than
$\kappa$ to those greater than $\kappa$, at any time $t$. This the well
known basis for the energy cascade.

It is usual to introduce a specific symbol $\Pi$ for this energy
flux, thus:
\beq
\Pi (\kappa,t) = \int^{\infty}_{\kappa} dk\, T(k,t) =-\int^{\kappa}_{0} dk\,
T(k,t),
\label{tp}
\eeq
where the second equality follows from (\ref{encon}).

\begin{figure} 
\begin{center}
\includegraphics[width=0.65\textwidth, trim=0px 200px 0px 200px,clip]{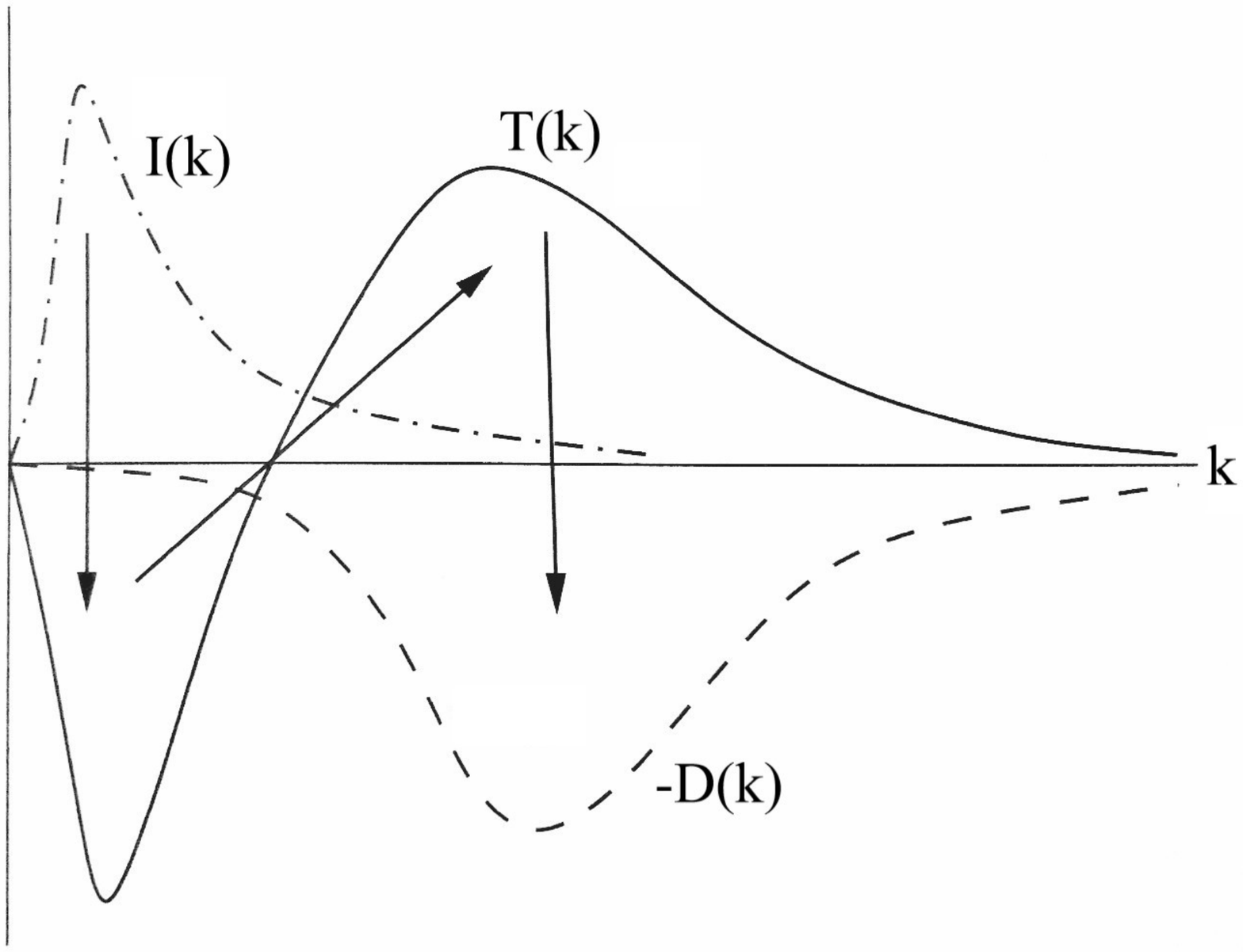}
\end{center} 
\caption{\small  A schematic view of the energy transfer in isotropic
turbulence. The input spectrum $I(k)$ can represent either the work spectrum
W(k) or $-\dd E(k.t)/ \dd t$; or the combined effects of both terms. All the
other symbols have their usual meaning as defined in the text.} 
\label{fig1} %In text, Figure \ref{figname} 
\end{figure}

In order to consider the stationary case, we may introduce an input
spectrum $W(k)$. It is also convenient to introduce the dissipation
spectrum $D(k,t)$ such that: 
\beq
D(k,t) = 2\nu k^2 E(k,t).
\eeq
With these introductions, and some rearrangement, we may write the
energy balance equation as:
\begin{equation}
\frac{\dd E(k,t)}{\dd t} = W(k) + T(k,t) - D(k,t).
\label{enbalt2}
\end{equation}
Figure (\ref{fig1}) illustrates the general form of the energy transfers involved.

%%GRAPH_FIG_CLIP%%%%%%%%%%%%%%%%%%%%%%%%%%%%%%%%%%%%%%%%%%%%%%%%%%
%%To insert a pdf figure with trimming of the image: parameters in
%%pixels (px)
%%Requires \usepackage{graphicx}
%% trim parameters <left> <bottom> <right> <top>
%%%%%%%%%%%%%%%%%%%%%%%%%%%%%%%%%%%%%%%%%%%%%%%%%%%%%%%%%%%%%%%%%%
It should be noted that this general schematic form applies both to the
stationary case and the case of free decay, with the input term $I(k)$
being interpreted as appropriate to each case.

\section{The paradox and its resolution}
%%%%%%%%%%%%%%%%%%%%%%%%%%%%%%%%%%%%%%%%%%%%%%%%%%%%%%%%%%

The inertial range of wavenumbers is defined as being where the time
derivative (or input term) and the viscous term  are negligible.
Hence, from equation (\ref{enbalt}), it
follows that the criterion for an inertial
range of wavenumbers can be taken as the vanishing of the transfer spectrum;
and, from equation (\ref{tp}),
the constancy of the flux. In other words, for wavenumbers $\kappa$ \emph{in
the inertial range} we might expect to have have:
\beq
T(\kappa,t)=0 \qquad \mbox{and} \qquad \Pi(\kappa,t) = \varepsilon.
\label{conflux}
\eeq

Scale invariance, can be summed up as the observation that the energy
spectrum takes the form of a power law (which is in itself scale-free)
and that there is a constant rate of energy transfer over a range of
wavenumbers, which must necessarily be equal to the rate of energy
dissipation. In practice, the second criterion of equation
(\ref{conflux}) is widely used to identify the inertial range. This
criterion was first put forward in 1941 by Obukhov \cite{Obukhov41} and
first used to derive the famous $-5/3$ spectrum using dimensional
analysis by Onsager in 1945 \cite{Onsager45}. More recently, the books
by Leslie \cite{Leslie73} and McComb \cite{McComb90a},\cite{McComb14a}
all follow Kraichnan \cite{Kraichnan59b}, and cite the criterion
$\Pi=\varepsilon$; as does  work by, for instance, Bowman
\cite{Bowman96}, Thacker \cite{Thacker97}, and Falkovich
\cite{Falkovich06}. However, the first criterion given in equation
(\ref{conflux}) only holds for a single wavenumber and this fact is the
scale-invariance paradox.

\begin{figure} 
\begin{center}
\includegraphics[width=0.65\textwidth, trim=0px 200px 0px 200px,clip]{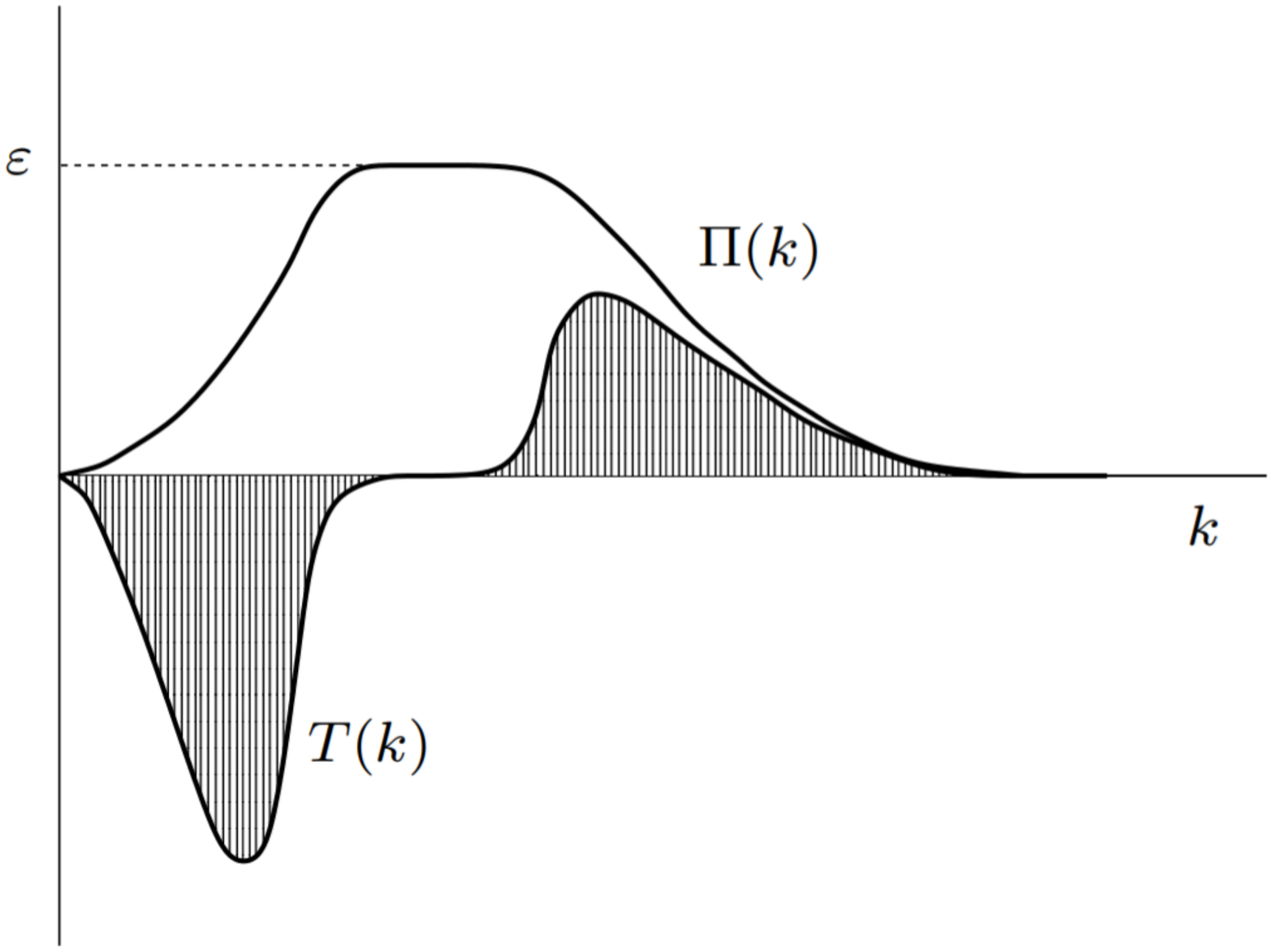}
\end{center} 
\caption{\small  The expected behaviour of $T(k)$, on the basis of
elementary calculus, to correspond to the scale invariance of $\Pi(k)$.
The fact that $T(k)$ does not behave like that is the scale-invariance paradox.} 
\label{fig2} %In text, Figure \ref{figname} 
\end{figure}

There are two inertial-range criteria in (\ref{conflux}); and, by
elementary calculus, they seem to be equivalent. This point is
illustrated in Fig. (\ref{fig2}). It shows an extended region where the
flux is constant and also the transfer spectrum is zero. This makes an
appealingly simple picture of spectral energy transfers but
unfortunately it is wrong. The transfer spectrum always passes through
zero at a single point as illustrated in Fig. (\ref{fig1}).

%Pict4
%%GRAPH_FIG_CLIP%%%%%%%%%%%%%%%%%%%%%%%%%%%%%%%%%%%%%%%%%%%%%%%%%%
%%To insert a pdf figure with trimming of the image: parameters in
%%pixels (px)
%%Requires \usepackage{graphicx}
%% trim parameters <left> <bottom> <right> <top>
%%%%%%%%%%%%%%%%%%%%%%%%%%%%%%%%%%%%%%%%%%%%%%%%%%%%%%%%%%%%%%%%%%

This property of $T(k)$ was first discovered in 1963 by Uberoi
\cite{Uberoi63} and later, extensive investigations confirmed that the
transfer spectrum always has a single zero-crossing
\cite{Bradshaw67,Helland77} and pragmatic, approximate  procedures were
introduced to allow the inertial range to be identified from the
behaviour of the transfer spectrum \cite{Lumley64}. For a discussion of
this topic, see \cite{McComb92}.

So, let us consider again equation (\ref{fluxbalt2}) for the transfer of
energy from low wavenumbers to high. Now we wish to draw attention to
the fact that, although the first term on the right hand side correctly 
represents the integral over wavenumber $k$ of the transfer spectrum
from zero up to $\kappa$, nevertheless the integrand is not actually
$T(k)$ (from now on, we shall suppress time arguments in the interests
of conciseness). In fact the integrand represents \emph{some part of}
$T(k)$, because the internal integration with respect to the dummy
variable $j$ has been truncated at $j=\kappa$. 

In order to clarify this situation, it will be found helpful to
introduce low- and high-pass filtering operations, based on a cut-off
wavenumber $k=\kappa$, on the Fourier components of the velocity field.
These operations are used for the study of spectral mode
elimination in the context of large-eddy simulation and its associated
subgrid modelling: see, for example, \cite{McComb01a} and
references therein. We are thus led to introduce transfer spectra which
have been filtered with respect to $k$ and which have had their
integration over $j$ partitioned at the filter cut-off, i.e. $j=\kappa$. 

Beginning with the Heaviside unit step function, defined by:
\bea
H(x) & = & 1 \qquad \mbox{for} \qquad x > 0; \\
& = & 0 \qquad \mbox {for} \qquad x < 0.
\eea
we may define low-pass and high-pass filter functions, thus:
\beq
\theta^{-}(x) = 1 - H(x),
\eeq
and
\beq
\theta^{+}(x) = H(x).
\eeq
We may then decompose the transfer spectrum, as given by (\ref{ts}), into four constituent parts, 
\beq
T^{--}(k|\kappa) = \theta^{-}(k-\kappa)\int^{\kappa}_{0}dj\, S(k,j);
\label{tmm}
\eeq

\beq
T^{-+}(k|\kappa) = \theta^{-}(k-\kappa)\int^{\infty}_{\kappa}dj\, S(k,j);
\label{tmp}
\eeq

\beq
T^{+-}(k|\kappa) = \theta^{+}(k-\kappa)\int^{\kappa}_{0}dj\, S(k,j);
\label{tpm}
\eeq
and
\beq
T^{++}(k|\kappa) = \theta^{+}(k-\kappa)\int^{\infty}_{\kappa}dj\, S(k,j),
\label{tpp}
\eeq
such that the overall requirement of energy conservation is satisfied:
\beq
\int^{\infty}_{0}dk\left[T^{--}(k|\kappa) +
T^{-+}(k|\kappa) + T^{+-}(k|\kappa) + T^{++}(k|\kappa)\right] = 0.
\eeq
It is readily verified that the individual filtered/partitioned transfer
spectra have the following properties:
\beq
\int^{\kappa}_{0}dk\, T^{--}(k|\kappa) = 0;
\label{mm}
\eeq

\beq
\int^{\kappa}_{0}dk\, T^{-+}(k|\kappa) = -\Pi(\kappa);
\label{mp}
\eeq

\beq
\int^{\infty}_{\kappa}dk\, T^{+-}(k|\kappa) = \Pi(\kappa);
\label{pm}
\eeq
and
\beq
\int^{\infty}_{\kappa}dk\, T^{++}(k|\kappa) = 0.
\label{pp}
\eeq

Equation (\ref{fluxbalt1}) may be rewritten in terms of the
filtered/partitioned transfer spectrum as:
\beq
\frac{d}{dt}\int^{\kappa}_{0}dk\, E(k,t) = -\int^{\infty}_{\kappa}dk\,
T^{+-}(k|\kappa) -2\nu_{0}\int^{\kappa}_{0}dk\, k^{2}E(k,t).
\label{fluxbaltmod}
\eeq
We note from equation (\ref{mm}) that $T^{--}(k|\kappa)$ is conservative on
the interval $[0,\kappa]$, and hence does not appear in
(\ref{fluxbaltmod}), while $T^{-+}(k|\kappa)$ has been replaced by
$-T^{+-}(k|\kappa)$, using (\ref{mp}) and (\ref{pm}).

Filtered and partitioned transfer spectra have been measured, using DNS,
in the context of spectral large-eddy simulation. In particular, Zhou
and Vahala \cite{Zhou93a} found  that the resolvable-scales energy
transfer spectrum $T^{<<}(k)$ (i.e. $T^{--}(k|\kappa)$ in our notation) is
conservative on the interval $0\leq k \leq \kappa$, in agreement with our
equation ({\ref{mm}}); while the resolvable-subgrid transfer spectrum
(i.e. our $T^{-+}(k|\kappa)$) is zero over a range of wavenumbers. Similar
behaviour has also been found in the more detailed investigation by
McComb and Young \cite{McComb98}.

%Pict5

%%GRAPH_FIG_CLIP%%%%%%%%%%%%%%%%%%%%%%%%%%%%%%%%%%%%%%%%%%%%%%%%%%
%%To insert a pdf figure with trimming of the image: parameters in
%%pixels (px)
%%Requires \usepackage{graphicx}
\begin{figure} 
\begin{center}
\includegraphics[width=0.65\textwidth, trim=0px 200px 0px 0px,clip]{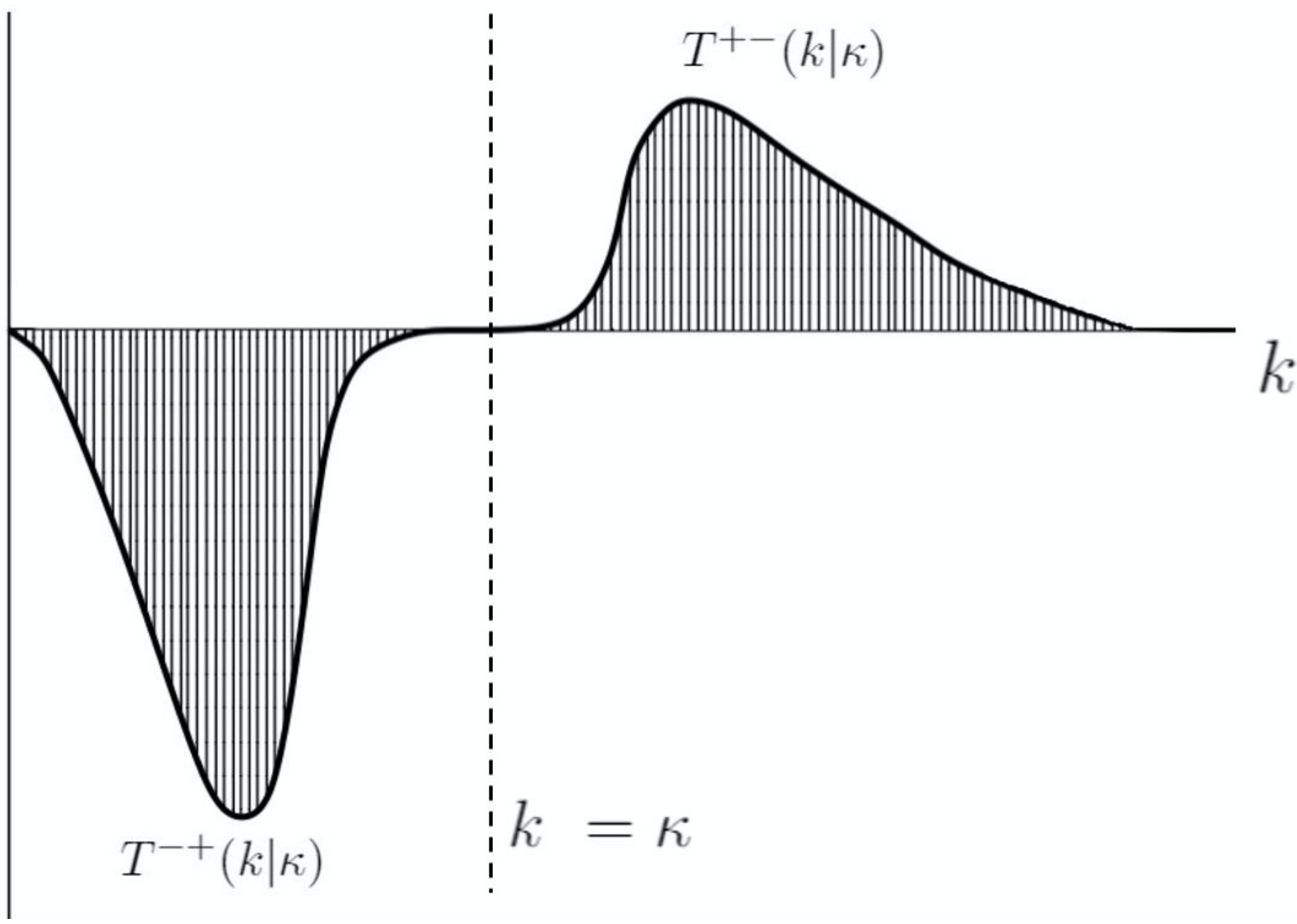}
\end{center} 
\caption{\small  The behaviour of the filtered-partitioned transfer
spectra: the paradox resolved!} 
\label{fig3} %In text, Figure \ref{figname} 
\end{figure}
%% trim parameters <left> <bottom> <right> <top>
%%%%%%%%%%%%%%%%%%%%%%%%%%%%%%%%%%%%%%%%%%%%%%%%%%%%%%%%%%%%%%%%%%

As we have previously pointed out in \cite{McComb08}, experimentalists,
who do not have access to partitioned versions of the transfer spectrum,
will still find pragmatic procedures, such as the Lumley criterion for
the inertial range \cite{Lumley64}, useful.  However, those working with
DNS or analytical theory, can avoid the paradox by changing their
definition of energy fluxes, from those given by (\ref{tp}), to the
forms\footnote{We should mention that these forms are exactly equivalent
to Kraichan's original definition of what he called the \emph{transport
power} \cite{Kraichnan59b}. In later work \cite{Kraichnan64b}, his
definition of the transport power was equivalent to equation (\ref{tp})
in the present paper.}: 

\beq \Pi (\kappa,t) = \int^{\infty}_{\kappa}
dk\, T^{+-}(k|\kappa,t) =-\int^{\kappa}_{0} dk\, T^{-+}(k|\kappa,t),
\label{tpmod} 
\eeq 
where $T^{+-}(k|\kappa,t)$ is defined by (\ref{tpm}) and
$T^{-+}(k|\kappa,t)$ by (\ref{tmp}). This is equivalent to (\ref{tp});
but, unlike it, avoids the paradox. This resolution of the paradox is
shown schematically in Fig. (\ref{fig3}).

\section{Modifications to the Lin equation}

In view of the above discussion, the obvious step now is to filter the
energy spectrum in the same way as we have done for the transfer
spectrum, and consider low-$k$ and high-$k$ forms of the Lin equation.
In order to do this we make the decomposition:
\beq
E(k,t) = E^-(k|\kappa,t) + E^+(k|\kappa,t),
\label{decompE}
\eeq
where $E^-$ is defined for $k\leq \kappa$ and $E^+$ is defined for
$k\geq \kappa$. Trivially, we can also do this for the input spectrum
$W(k)$ and dissipation spectrum $D(k,t)$, and equation (\ref{enbalt2})
can be written in low-$k$ and high-$k$ forms respectively, as:
\begin{equation}
\frac{\dd E^-(k|\kappa,t)}{\dd t} = W^-(k|\kappa) + T^{--}(k|\kappa,t) +
T^{-+}(k|\kappa,t) - D^-(k|\kappa,t), \quad \mbox{for}\quad k \leq \kappa;
\label{enbalt_low}
\end{equation} 
and
\begin{equation}
\frac{\dd E^+(k|\kappa,t)}{\dd t} = W^+(k|\kappa) + T^{+-}(k|\kappa,t) +
T^{++}(k|\kappa,t) - D^+(k|\kappa,t), \quad \mbox{for}\quad k \geq \kappa.
\label{enbalt_high}
\end{equation}
For this decomposition to be meaningful, the Reynolds number must be
large enough for the inertial flux to be equal to the dissipation, in
accordance with 
the second criterion of equation (\ref{conflux}). As we increase the
Reynolds number beyond this critical value, we have an increasing range
of wavenumbers $k$ which satisfy that criterion, and this is the
\emph{inertial range}. We shall denote this range by 
\[
k_{\bote} \leq k \leq k_{\tope} \quad \equiv \quad \mbox{the inertial
range of wavenumbers,}
\]
where we now have to define $k_{\bote}$ and $k_{\tope}$. For sake of
simplicity, we will consider stationary turbulence and omit the time
variables. 

First, we need to consider the nature of the forcing spectrum $W(k)$. In
formulating the turbulence problem according to the tenets of
statistical physics, this is normally taken to arise from the
introduction of random stirring forces, which are assumed to be of
\emph{white noise} form. In particular, the forcing spectrum is taken to
be peaked near the origin in wavenumber space, so that the turbulence
that results from it is due to the Navier-Stokes equation, and not
specifically related to the forcing. We should note that a
different view was taken from the late 1970s onwards, in connection with
the application of renormalization group methods to the Navier-Stokes
equation. See either of the books \cite{McComb90a} or \cite{McComb14a}
for a general discussion of this point.

Accordingly, for theoretical approaches to the statistical closure
problem, and also for direct numerical simulation, we should choose a
form of forcing spectrum $W(k)$ which satisfies the conditions:
\beq
\int_0^\infty dk W(k) = \varepsilon_W \simeq \int_0^{k_{\bote}} dk W(k),
\label{kbot}
\eeq
where the equality defines $\varepsilon_W$, while the approximate
equality defines $k_{\bote}$, which we take to be the lower limit of the
inertial range.

In general, we would require $k_{\bote}$ to be very much smaller than
the Kolmogorov dissipation wavenumber $k_d$ which is generally taken as
being an indicator of the dissipation range of wavenumbers.
Experimenters have usually taken the the upper limit of the
inertial range to be about $0.1 k_d - 0.2 k_d$. In fact we will define
$k_{\tope}$ by another approximate equality, thus:
\beq
\int_0^\infty dk D(k) = \varepsilon \simeq \int_{k_{\tope}}^\infty dk D(k),
\label{ktop}
\eeq
where the equality is the conventional definition of the dissipation
rate, and the approximate equality defines the upper limit of the
inertial range $k_{\tope}$.

With these points in mind, we may simplifly the low-wavenumber and
high-wavenumber forms of the Lin equation, respectively
(\ref{enbalt_low}) and (\ref{enbalt_high}), to:
\begin{equation}
\frac{\dd E^-(k|\kappa,t)}{\dd t} = W(k) + T^{--}(k|\kappa,t) +
T^{-+}(k|\kappa,t), \quad \mbox{for}\quad k \leq \kappa;
\label{Lin-low}
\end{equation} 
and
\begin{equation}
\frac{\dd E^+(k|\kappa,t)}{\dd t} = T^{+-}(k|\kappa,t) +
T^{++}(k|\kappa,t) - D(k,t), \quad \mbox{for}\quad k \geq \kappa.
\label{Lin-high}
\end{equation}
That is, for sufficiently high Reynolds numbers, and an appropriate
choice of stirring forces, we may simplify matters by treating the input
spectrum as being confined to the low-wavenumber region and the
dissipation spectrum  as being confined to the high-wavenumber region.
Deriving the flux balance equations from (\ref{Lin-low}) and
(\ref{Lin-high}), and invoking equations (\ref{kbot}) and (\ref{ktop}),
we obtain the final flux balances as:
\beq
\varepsilon_W - \Pi(\kappa) = 0 \quad \mbox{for}\quad k \leq \kappa; 
\eeq
and
\beq
\Pi(\kappa) - \varepsilon = 0 \quad \mbox{for}\quad k \geq \kappa.
\eeq

Reminding ourselves that the transfer spectrum has its single zero
crossing at $k=k_\ast$, we may define the maximum value of the inertial
flux as
\beq
\Pi_{\mbox{max}} = \Pi(k_\ast) = \varepsilon_T,
\eeq
and at the same time introduce the useful symbol $\varepsilon_T$ for the
maximum flux. Since $k_\ast$ must lie within the inertial range, we can
write the general criterion for the existence of the inertial range as:
\beq
\Pi(\kappa) = \varepsilon_T = \varepsilon_W = \varepsilon.
\eeq
For completeness it should be noted that this analysis is readily
extended to the case of free decay, if we replace $\varepsilon_W$ by the
energy decay rate $\varepsilon_D$. Further details may be found in
\cite{McComb14a}. 

\section{Conclusion}

Provided we are faced with the ideal situation, where the input and the
output (\emph{i.e.} dissipation) are well separated in wavenumber space,
equations (\ref{Lin-low}) and (\ref{Lin-high}) may provide a new, and one
might hope, productive basis for the study of the energy transfers in
isotropic turbulence. The corresponding partitioned-filtered
Navier-Stokes equations are readily deduced and may be studied by direct
numerical simulation as a four-component composite dynamical system,
where the four components correspond to the four filtered-partitioned
transfer spectra. 

Also, there is a growing use of hybrid approaches in fluid dynamics
problems, and the closure problem could be approached in such a way by
using different methods to tackle the different filtered-partitioned
transfer spectra. For instance, in the low-$k$ system, we might use the
local energy transfer theory \cite{McComb17a} for $T^{--}(k)$, and
renormalization group methods \cite{McComb06} for $T^{-+}(k)$; or,
conceivably, the other way round! It would require investigation. 

For the ideal situation just discussed, where we have the input and
output (or, production and  dissipation) ranges of wavenumber well
separated, we need to choose the input spectrum $W(k)$ to be peaked near
the origin; and also we need the Reynolds number to be reasonably high.
If for some reason, we cannot satisfy these conditions, then we must
resort to equations (\ref{enbalt_low}) and (\ref{enbalt_high}). However,
even so, we must still have the Reynolds number large enough for the
condition for the existence of an inertial range to be satisfied.

Lastly, I should emphasise that Fig. (\ref{fig3}) is very much a
schematic indication of how this graph should look, based on the small
amount of information available to us. The behaviour of these
filtered-partitioned transfer spectra was studied in the 1990s in the
context of subgrid modelling and renormalization group methods: see
\cite{McComb08} for references. Computers have advanced a lot since
then, so we end with a plea to the effect that this field of study
should be revived in the context of later work. An informal introduction
to this topic may be found in the post of 23 July on the following
weblog: blogs.ed.ac.uk/physics-of-turbulence/.

\section*{Acknowledgements}

I wish to thank John Morgan who worked on this topic with me as part of
his MPhys research project in the academic year 2018/19. It was John's
idea to plot Fig. (\ref{fig3}) in order to make the resolution of the
scale-invariance paradox clearer and he also prepared the figures.

% to use bibtex with e.g. wdm.bib: latex article, bibtex article,
% latex(x2) article
%\bibliographystyle{unsrt}
%\bibliography{wdm}

\end{document}